\documentclass[usegraphicx,usenatbib,onecolumn,doublespace]{mn2e}
\usepackage{amsmath}
\usepackage{amssymb}
\usepackage{bm}
\begin{document}
\title{Second order cross-correlation between kSZ and 21 cm fluctuations 
from the EoR} 
\author[Tashiro, H. et al.]
{Hiroyuki Tashiro$^{1,2}$, Nabila Aghanim$^{2,3}$, Mathieu Langer$^{2,3}$,
\newauthor Marian Douspis$^{2,3}$, Saleem Zaroubi$^{4,5}$, and Vibor Jeli{\'c}$^{6}$ \\
$^1$ Center for Cosmology, Particle Physics and Phenomenology (CP3),\\   Univ. catholique de Louvain,%
B-1348 Louvain-la-Neuve, Belgium;\\
$^2$ Univ. Paris-Sud, Institut d'Astrophysique Spatiale, UMR6817, Orsay, F-91405, France;\\
$^3$ CNRS, Orsay, F-91405, France;\\
$^4$ Kapteyn Astronomical Institute, University of Groningen, P.O. Box 800, NL-9700AV, Groningen, The Netherlands;\\
$^5$ Physics Department, Technion, Haifa 32000, Israel;\\
$^6$ ASTRON, P.O. Box 2, NL-7990AA, Dwingeloo, the Netherlands
}

\date{\today}
\maketitle

\begin{abstract}
The measurement of the brightness temperature fluctuations of 
neutral hydrogen 21 cm lines from the Epoch of Reionisation (EoR) is
expected to be a powerful tool for revealing the reionisation 
process. We study the 21 cm cross-correlation with Cosmic Microwave
Background (CMB) temperature anisotropies, focusing on the effect of
the patchy reionisation. We calculate, up to second order, the angular power spectrum 
of the cross-correlation between 21 cm fluctuations and the CMB 
kinetic Sunyaev-Zel'dovich effect (kSZ) from the EoR, using an analytical 
reionisation model.
We show that the kSZ and the 21 cm fluctuations
are anti-correlated on the scale corresponding to the typical 
size of an ionised bubble at the observed redshift of the 21 cm fluctuations. The
amplitude of the angular power spectrum of the cross-correlation 
depends on the fluctuations of the ionised fraction. Especially, in 
a highly inhomogeneous reionisation model, the amplitude reaches the
order of $100~\mu{\rm K}^2$ at $\ell \sim 3000$.  
We also show that second order terms may help in distinguishing between reionisation histories. 
\end{abstract}


\maketitle

\section{Introduction}

The Epoch of Reionisation (EoR) is an essential milestone in the formation and evolution of cosmic structure.
The first luminous objects produced in collapsed dark matter halos 
in the early universe ($z \sim 20$)  started to reionise the inter galactic medium (IGM) which
was neutral after recombination. Currently we have only 
a few observations for the EoR. 
The first one is the Ly-$\alpha$ absorption measurement 
towards high redshift QSOs which probes the fraction of neutral hydrogen along the line of sight
\citep{2006AJ....132..117F},
and the second one is the large-scale CMB polarisation \citep{2010arXiv1001.4538K}. 
These observations indicate that the
IGM was fully ionised by redshift $z \sim 6$.
The recent HST observations found large samples of Lyman break 
galaxies (LBGs) at high redshifts, 
$7\lesssim z \lesssim 10$ \citep{2010arXiv1006.4360B}. \citet{2010arXiv1006.4360B} have studied  reionisation 
with a galaxy model based on these data. Their results suggested that, in addition to such high redshift LBGs, other
reionisation sources, for example, faint galaxies and population III
stars, are required to match the optical depth of the WMAP seven-year data.

While, current observational data for the EoR are insufficient to study the details of the EoR.
In recent years, several observations of signals from the EoR have been suggested
to obtain further information about the EoR,
for example fluctuations of the neutral hydrogen 21~cm line
(\citealt{madau-meiksin-rees-1997}, for a review see \citealt{2006PhR...433..181F}),
small-scale CMB anisotropies due to the kinetic Sunyaev-Zel'dovich (kSZ; \citealt{1980MNRAS.190..413S,1986ApJ...306L..51O,1987ApJ...322..597V},
for a review see \citealt{2008RPPh...71f6902A}), and 
Ly-$\alpha$ damping of high redshift QSOs and gamma ray bursts
\citep{1998ApJ...501...15M,2004ApJ...601...64B}. 
While the latter can provide us with information about 
the end of the EoR,
the former two are expected to probe the IGM during the EoR. The
LOFAR\footnote{http://www.lofar.org}, MWA\footnote{http://www.mwatelescope.org/}
and SKA\footnote{http://www.skatelescope.org} are being installed or designed for the measurement of
 21~cm line fluctuations, while telescopes such as
ACT\footnote{http://www.physics.princeton.edu/act/}, SPT\footnote{http://pole.uchicago.edu/} and OLIMPO \citep{2008MmSAI..79..887M}
will be used to detect and measure the kSZ signal.

Although both auto-correlations of 21~cm lines and CMB anisotropies during the EoR are good probes of the EoR,
the cross-correlation between 21~cm fluctuations and the CMB anisotropies 
created during the EoR is also expected to be useful to study the history of the EoR.
The cross-correlation has a potential to provide additional information other than 
their respective auto-correlations.  
Besides, the cross-correlation decreases the statistic errors caused by
the foreground and the systematic effects, as compared to their auto-correlation.  
There are several analytical or numerical works about the cross-correlation
between CMB and 21 cm fluctuations during the EoR.
\citet{2006ApJ...647..840A} and \citet{2008MNRAS.384..291A}
computed the expected signal on large scales ($\ell \sim 100$)
by analytically calculating the cross-correlation between 21 cm fluctuations
and the CMB Doppler anisotropies in the linear regime of the cosmological perturbations.
\citet{2010MNRAS.402.2617T} studied the detectability of these signals by
LOFAR, MWA and SKA. On small scales ($\ell >1000$), because
the dominant contributions of CMB anisotropies come 
from the kSZ effect due to the patchiness of the ionised medium,
\citet{2004PhRvD..70f3509C} has partially studied 
the cross-correlation with kSZ anisotropies
and the second order 21 cm fluctuations in a simple reionisation model. 
%And \citet{2004PhRvD..70f3509C} 
He has also investigated the higher order
cross-correlation by calculating the bispectrum.
\citet{2007MNRAS.377..168S} have also done the study of the
21~cm cross-correlation with the CMB SZ effect
which is caused by hot electrons in the first
supernovae remnants during the EoR.
Since reionisation is a complex physical process, numerical simulations play an important role
 in the studies of the 21~cm cross-correlation
with CMB temperature anisotropies. Numerical works
by \citet{2010MNRAS.tmp...65J} and \citet{2005MNRAS.360.1063S},
focus especially on the small-scale cross-correlation due to the patchy reionisation.
Additionally, the 21~cm cross-correlation with CMB polarisation has been calculated
by \citet{2008MNRAS.389..469T} and \citet{2009PhRvD..79j7302D}.
%ISW-21cm \citep{2009JCAP...08..019G}.

In this paper, we study the cross-correlation 
between kSZ anisotropies and the second order 21 cm fluctuations during the EoR
analytically. 
%The second order CMB temperature anisotropies are produced by
%the kinetic SZ effect from patchy ionised medium.
\citet{2004PhRvD..70f3509C} has studied this cross-correlation
in the simple analytical reionisation model where the fluctuations of the ionisation fraction
are linearly related to the density fluctuations. He concluded that the cross-correlation cannot appear
due to the geometric cancellation occurring between 
the velocity and the density fluctuations.
However, the kSZ effect depends strongly on the evolution of the ionisation bubbles, 
and numerical studies of the cross-correlation between 21~cm and kSZ anisotropies 
also shows that patchy reionisation generates signals
 on small scales
\citep{2005MNRAS.360.1063S}.
Therefore, we revisit this issue with the analytical model
of \citet{2005ApJ...630..643M} which produces a reionisation 
history similar to that found in recent numerical simulations.

The outline of our paper is the following.
In Sec.~II, we give the analytical form of the second order
cross-correlation between kSZ anisotropies and
 21~cm fluctuations.
In Sec.~III, we give a short description of the analytical reionisation model
based on \citet{2005ApJ...630..643M}.
In Sec.~IV, we show the angular power spectrum of the second order cross-correlation
and we discuss the detectability in the case of the SKA sensitivity.
Section V is devoted to the conclusions.  
Throughout the paper, we use the concordance cosmological parameters for a
flat cosmological model, i.e. $h=0.73 \ (H_0=h
\times 100 {\rm ~km/s / Mpc})$, $T_0 = 2.725$K, $\Omega _{\rm b}
=0.05$, $\Omega_{\rm m} =0.27$ and $\sigma_8=0.9$.

\section{The second order cross-correlation}

In this section, we calculate the angular power spectrum 
of the cross-correlation
between 21~cm fluctuations and kSZ anisotropies
during the EoR at the second order in the fluctuations.
For simplicity, we assume that both fluctuation fields are 
isotropic statistically. Under this assumption,
the angular power spectrum of the cross-correlation 
$C_\ell$ is given by
\begin{equation}
\langle a_{\ell_1 m_1} ^{* {\rm kSZ} } a_{\ell_2 m_2}^{21}\rangle
=\delta_{\ell_1 \ell_2} ^D \delta_{m_1 m_2} ^D 
C_{\ell_1}, 
\label{eq:defcrosscl}
\end{equation}
where $a_{\ell_1 m_1}^{{\rm kSZ}}$ and $a_{\ell_2 m_2}^{21}$
are the multipole components of the CMB temperature anisotropies
and 21~cm fluctuations during the EoR.

\subsection{kSZ CMB anisotropies}

During the EoR, secondary CMB temperature anisotropies are caused by 
the kinetic SZ effect. Their expression is 
%The produced temperature fluctuations are calculated by
\begin{equation}
T_{\rm kSZ} (\hat {\bm n})=-T_{\rm cmb} \int^{\eta_0} ~d\eta 
~ g(\eta) \hat {\bm n} \cdot  {\bm v} (\eta, \hat {\bm n}),
\label{eq:ksz}
\end{equation}
where ${\bm v}$ is the baryon velocity field, 
$g(\eta)$ is the visibility function at 
the conformal time $\eta$, and the present value of the conformal 
time is $\eta_0$.
The visibility function is given by
$g(\eta) = \dot \tau e^{-\tau}$ where
$\tau$ is the optical depth of Thomson scattering
from $\eta$ to today and
$\dot \tau = \sigma_T x_i n_{\rm H}$ 
with $\sigma_T$ the cross section of Thomson scattering,
$x_i$ the ionised fraction, and $n_{\rm H}$ the neutral
hydrogen density (we ignore the ionisation of helium).

We can decompose $x_i$ and $n_{\rm H}$
into the background and fluctuation values,
\begin{equation}
n_{H}=\bar n_{H} (1 +\delta), \qquad
x_i=\bar x_i (1+\delta_x),
\label{eq:flucdeco}
\end{equation}
where the symbols with a bar represent the background values.
In Eq.~(\ref{eq:flucdeco}), since we assume that the hydrogen density 
follows the dark matter density on scales much bigger than the baryonic Jeans length,
 $\delta$ is the total matter density fluctuation field.

Substituting Eq.~(\ref{eq:flucdeco}) into Eq.~(\ref{eq:ksz}),
we obtain
\begin{equation}
T_{\rm kSZ} (\hat {\bm n})=-T_{\rm cmb} \int ~d\eta ~ {\bar g} (\eta)
\hat {\bm n} \cdot  {\bm v} (\eta, \hat {\bm n})
~(1+\delta (\eta, \hat {\bm n})+ \delta_x (\eta, \hat {\bm n})
+\delta (\eta, \hat {\bm n}) \delta_x (\eta, \hat {\bm n})).
\label{eq:ksz2}
\end{equation}

We focus on the second order part in Eq.~(\ref{eq:ksz2}),
which can be written  in terms of the Fourier components of the fluctuations as
\begin{equation}
\delta T_{\rm kSZ} (\hat {\bm n})= -iT_{\rm cmb} \int d\eta \int {d^3 {\bm k} \over (2 \pi)^3} 
\int{d^3 {\bm k'}  \over (2 \pi)^3} {\bar g} (\eta)
\frac{\hat {\bm n} \cdot ({\bm k}-{\bm k}')}{|{\bm k}-{\bm k}'|^2} 
\dot \delta (\eta,{\bm k}-{\bm k}')
(\delta (\eta, {\bm k}')+ \delta_x (\eta, {\bm k}'))
\exp[i (\eta_0-\eta) (\hat{\bm n} \cdot {\bm k})],
\label{eq:fourierksz}
\end{equation}
where we use the relation ${\bm r}= (\eta_0-\eta) \hat  {\bm n }$,
and we relate the velocity to $\delta$ by the continuity equation in 
the cosmological linear perturbation theory
\begin{equation}
{\bm v} = i \frac{\bm k}{k^2} \dot \delta (\eta, k),
\end{equation}
where the dot represents the derivative with respect to $\eta$.

Our final aim is to obtain the angular power spectrum of the cross-correlation.
Therefore, we consider the spherical harmonic decomposition of Eq.~(\ref{eq:fourierksz}),
 $a_{\ell m} ^{\rm kSZ}=\int d \hat {\bm n} \delta T_{\rm kSZ} (\hat {\bm n}) Y_{\ell} ^m$.
The spherical harmonic coefficients of the kSZ are given by
\begin{eqnarray}
a_{\ell m}^{\rm kSZ}
&=&\sum_{\substack{\ell' m'  \ell'' m'' \\ \ell''' m'''  m''''}}
\int d\eta 
\int {d^3 {\bm k}_1 \over (2 \pi)^3} 
\int {d^3 {\bm k}_2  \over (2 \pi)^3}
\nonumber \\
&&\quad \times
A_{\ell \ell'  \ell'' \ell''' 1 } ^{m m'  m'' m''' m''''} (\eta)
\dot \delta (\eta,{\bm k}_1)
(\delta (\eta, {\bm k}_2)+\delta_{x}(\eta, {\bm k}_2) )
{j_{\ell'}(k_1 r) \over k_1} j_{\ell''}(k_2 r)
Y^{m'''}_{\ell'''}( \hat {\bm k}_1)
Y_{\ell''}^{m''}( \hat {\bm k}_2),
\label{eq:multi-ksz}
\end{eqnarray}
where we replaced ${\bm k}$ and ${\bm k}'$ by
${\bm k}_1 \equiv {\bm k} - {\bm k}'$ and ${\bm k}_2 \equiv {\bm k}'$, and
\begin{equation}
A_{\ell \ell'  \ell'' \ell''' 1  } ^{m m'  m'' m''' m''''}
=
-i  
(-1)^{m+m'-m''+m''''} {64 \pi^3 \over 3 } i^{\ell' + \ell''}
 \sqrt{ 3(2 \ell' +1)  \over 4 \pi (2 \ell'''' +1)}  
C^{\ell' \ell''' 1} _{-m' -m''' m''''}
C^{\ell' \ell''' 1} _{000}
M^{-m m' -m'' m''''}_{\ell \ell' \ell'' 1} T_{\rm cmb} \bar g(\eta).
\end{equation}
Here, $C^{\ell_1 \ell_2  \ell} _{m_1 m_2 m}$ are the Clebsch-Gordan
coefficients and $M^{m_1 m_2  m_3  m_4 } _{\ell_1 \ell_2  \ell_3 \ell_4}$
are the integrals of quadruple spherical harmonics,
\begin{eqnarray}
M^{m_1 m_2  m_3  m_4 } _{\ell_1 \ell_2  \ell_3 \ell_4} 
& = &
\int d\hat n~
Y^{m_1 }_{\ell_1} (\hat n) Y^{m_2 }_{\ell_2} (\hat n) Y^{m_3 }_{\ell_3} (\hat n) Y^{m_4 }_{\ell_4} (\hat n)
\nonumber \\
& = &
(-1)^{m_1}
\sum_{\ell' m'} 
\sqrt{  (2 \ell _2 +1)( 2 \ell _3 +1) (2 \ell_4 +1) \over 16 \pi^2 (2 \ell_1 +1)}
C^{\ell_3 \ell_4  \ell'} _{m_3 m_4 m'}
C^{\ell_3 \ell_4  \ell'} _{000}
C^{\ell_2 \ell'  \ell_1} _{m_2 m' -m_1}
C^{\ell_2 \ell'  \ell_1} _{000}.
\end{eqnarray}
%({\it The detail calculation is shown in Sec.~\ref{sec:kszcal}.})

\subsection{21~cm fluctuations}

%The observed brightness temperature of the 21~cm lines in a direction
%$\hat {\bm n}$ and at a frequency $\nu$ is given as in
%\cite{madau-meiksin-rees-1997} by
%\begin{equation}
%T_{21} (\hat {\bm n},\nu) = \frac{\tau_{21}}{(1+z_{\rm obs})}
%(T_{\rm s} -T_{\rm CMB})(z_{\rm obs}, \hat {\bm n}),
%% (\eta_0-\eta_{\rm obs})
%\label{eq:21cmline}
%\end{equation}
%where $T_{\rm CMB}$ is the CMB temperature and  $T_{\rm s}$
%is the spin temperature given by the ratio of the number
%density of hydrogen in the excited state to that of hydrogen in the
%ground state.  The redshift $z_{\rm
%obs}$ is related to the frequency $\nu$ as $\nu = \nu_{21}/(1+z_{\rm obs})$ with 
%$\nu_{21}$ being the frequency corresponding to the 21~cm wavelength.

The brightness temperature of the 21~cm line from a redshift $z$
is given as in \cite{madau-meiksin-rees-1997} by
\begin{equation}
T_{21} (z) = \frac{\tau_{21}}{(1+z)}
(T_{\rm s} -T_{\rm CMB})(z),
\label{eq:21cmline}
\end{equation}
where $T_{\rm CMB}$ is the CMB temperature and  $T_{\rm s}$
is the spin temperature given by the ratio of the number
density of hydrogen in the excited state to that of hydrogen in the
ground state.  
%The redshift $z_{\rm
%obs}$ is related to the frequency $\nu$ as $\nu = \nu_{21}/(1+z_{\rm obs})$ with 
%$\nu_{21}$ being the frequency corresponding to the 21~cm wavelength. 
The optical depth for the 21~cm line absorption $\tau_{21}$ is 
%\citep{bharadwaj-ali-2004}
\begin{equation}
\tau_{21} (z)
= {3 c^3 \hbar A_{10} x_{\rm H} n_{\rm H} \over 16 k \nu_{21} ^2 
T_{\rm s} H(z)}
%\left[ 1-{(1+z) \over H(z)} {\partial v \over \partial r}
%\sim 8.6 \times 10^{-3} x_{\rm H} \frac{T_{\rm cmb}}{T_{\rm s}}
%\left ( \frac{\Omega h^2}{0.02} \right)
%\left[
%\left ( \frac{0.15}{\Omega_{\rm m} h^2} \right)
%\left ( \frac{1+z_{\rm obs}} {10} \right)
%\right] ^{1/2}
,
\label{eq:tau21}
\end{equation}
where $A_{10}$ is the Einstein A-coefficient, $\nu_{21}$ is the
frequency corresponding to the 21~cm wavelength and
%$v$ is the line-of-sight component of the peculiar velocity and 
$x_{\rm H}$ is the fraction of neutral hydrogen, which is written as a
function of the ionised fraction $x_i = 1- x_{\rm H}$.
Note that we drop the redshift space distortion by the peculiar velocity
fluctuations of neutral hydrogen in Eq.~({\ref{eq:tau21}), 
although this effect enhances the 21~cm fluctuations \citep{bharadwaj-ali-2004}.

%Combining Eq.~(\ref{eq:flucdeco}) with Eqs. (\ref{eq:21cmline}) and (\ref{eq:tau21}),
%we can decompose the 21~cm signals into the background and the fluctuation parts.
%The second order fluctuations which we focus is given by
%\begin{equation}
%\delta T_{21} (\hat {\bm n}, \nu)=  \int ~d\eta \int {d^3 {\bm k} \over (2 \pi)^3} 
%{d^3 {\bm k'}  \over (2 \pi)^3}
%W_{21}(\eta, \eta(z_{\rm obs})) T_{21}(z(\eta)) 
%\delta (\eta,{\bm k}-{\bm k}')
%\delta_{H} (\eta, {\bm k}') \exp[i (\eta_0-\eta) (\hat{\bm n} \cdot {\bm k})],
%\end{equation}
%where $W_{21}(\eta, \eta(z))$ is a spectral response function of the 
%observation experiment which is normalised as $\int d \eta W_{21}(\eta, \eta(z)) =1$ 
%and centered at $\eta(z)$,
%$\delta _{H} \equiv (x_H - \bar x_H)/\bar x_H$
%and  $T_{21}$ is a normalisation temperature factor given by 
%\begin{equation}
%T_{21}(z) =23  \left({\Omega_{\rm b} h^2 \over 0.02} \right)
%\left[ \left({0.15\over \Omega_{\rm m} h^2} \right)  
%\left( {1+z \over 10} \right) \right] ^{1/2} \left({T_{\rm s}-T_{\rm cmb} \over T_{\rm s}} \right)~{\rm mK}. 
%%T_0 =23  ({\Omega h^2})/({0.02} )
%%[({0.15})/({\Omega_{\rm m} h^2} )  ( {1+z_{\rm
%%      obs}})/({10} ) ] ^{1/2} (T_{\rm s}-T_{CMB})/T_{\rm s}{\rm mK}. 
%\end{equation}

Combining Eq.~(\ref{eq:flucdeco}) with Eqs. (\ref{eq:21cmline}) and (\ref{eq:tau21}),
we can obtain the observed 21~cm fluctuations at the observed
frequency $\nu$. The second order fluctuations which we here focus on is given by
\begin{equation}
\delta T_{21} (\hat {\bm n}, \nu)=  \int ~d\eta \int {d^3 {\bm k} \over (2 \pi)^3} 
{d^3 {\bm k'}  \over (2 \pi)^3}
W_{21}(\eta, \eta(z_{\rm obs})) T_{0}(z(\eta)) 
\delta (\eta,{\bm k}-{\bm k}')
\delta_{H} (\eta, {\bm k}') \exp[i (\eta_0-\eta) (\hat{\bm n} \cdot {\bm k})],
\end{equation}
where $W_{21}(\eta, \eta(z))$ is the spectral response function of the 
observation experiment, normalised as $\int d \eta W_{21}(\eta, \eta(z)) =1$ 
and centred at $\eta(z)$,
the redshift $z_{\rm obs}$ is related to the frequency 
$\nu$ as $\nu = \nu_{21}/(1+z_{\rm obs})$,
$\delta _{H} \equiv (x_H - \bar x_H)/\bar x_H$
and  $T_{0}$ is a normalisation temperature factor given by 
\begin{equation}\label{eq:tzero}
T_{0}(z) =23  \left({\Omega_{\rm b} h^2 \over 0.02} \right)
\left[ \left({0.15\over \Omega_{\rm m} h^2} \right)  
\left( {1+z \over 10} \right) \right] ^{1/2} \left({T_{\rm s}-T_{\rm cmb} \over T_{\rm s}} \right)~{\rm mK}. 
%T_0 =23  ({\Omega h^2})/({0.02} )
%[({0.15})/({\Omega_{\rm m} h^2} )  ( {1+z_{\rm
%      obs}})/({10} ) ] ^{1/2} (T_{\rm s}-T_{CMB})/T_{\rm s}
\end{equation}
The spin temperature is determined by three couplings with CMB, IGM gas and Ly-$\alpha$ photons.
In the EoR, Ly-$\alpha$ photons emitted from ionising sources couple 
the spin temperature with the IGM gas temperature \citep{ciadri-madau-2003}.
Meanwhile, since the IGM gas is heated up quickly 
by Ly-$\alpha$ and X-ray photons from stars and QSOs, the IGM gas temperature is much 
higher than the CMB temperature during reionisation. Therefore, we can assume 
$T_{\rm s} \gg T_{\rm cmb}$ during the EoR in Eq. (\ref{eq:tzero}). 

Taking the harmonic decomposition, we obtain the
spherical harmonic coefficients of the 21~cm fluctuations,
\begin{equation}
a_{\ell m}^{21} = 
\sum_{\ell' m'} \sum_{\ell'' m''} 
\int ~d\eta 
\int {d^3 {\bm k _1} \over (2 \pi)^3} 
\int{d^3 {\bm k _2}  \over (2 \pi)^3}
B^{m m' m''}(\eta)
 \delta (\eta,{\bm k}_1)
\delta_{H} (\eta, {\bm k}_2) 
j_{\ell'}(k_1 r) j_{\ell''}(k_2 r) 
Y_{\ell'}^{m'*}(\hat {\bm k}_1) Y_{\ell''}^{m''*}( \hat {\bm k}_2),
\label{eq:multi-21cm}
\end{equation}
where
\begin{equation}
B^{m m' m''}(\eta)
=
16 \pi^2 i^{\ell' + \ell''} W_{21}(\eta) T_{0}(\eta)
M^{m' m'' -m}_{\ell' \ell'' \ell}.
\end{equation}
Here $M^{m_1 m_2 m_3} _{\ell_1 \ell_2 \ell_3}$ is the integral of triple spherical harmonics,
\begin{eqnarray}
M^{m_1 m_2 m_3} _{\ell_1 \ell_2 \ell_3} 
& = &
 \int d\hat n~
Y^{m_1}_{\ell_1} (\hat n) Y^{m_2}_{\ell_2} (\hat n) Y^{m_3}_{\ell_3} (\hat n)
\nonumber \\ 
&=&
(-1)^{m_1} 
\sqrt{ (2 \ell _2 +1)( 2 \ell _3 +1) \over 4 \pi (2 \ell_1 +1)}
C^{\ell_2 \ell_3  \ell_1} _{m_2 m_3 -m_1}
C^{\ell_2 \ell_3  \ell_1} _{000},
\end{eqnarray}
where $m_1+m_2=m_3$.

%({\it The detail calculation is given in Sec.~\ref{sec:21cmcal}.})

\subsection{The cross-correlation}

The second order cross-correlation is given by
substituting Eqs.~(\ref{eq:multi-ksz}) and (\ref{eq:multi-21cm})
into Eq.~(\ref{eq:defcrosscl}).
We obtain 
\begin{eqnarray}
C^{{\rm kSZ}-21} _{\ell}
&=&-
\sum_{\ell'_1 m'_1} \sum_{\ell''_1 m''_1} 
\sum_{\substack{\ell'_2 m'_2 \\ \ell''_2 m''_2 m'''_2 \\ \ell''''_2 m''''_2}}
\int d\eta 
\int d\eta'
\int {d^3 {\bm k_1} \over (2 \pi)^3} 
\int {d^3 {\bm k_2} \over (2 \pi)^3}
\int {d^3 {\bm k_1'} \over (2 \pi)^3} 
\int {d^3 {\bm k_2'} \over (2 \pi)^3}
\nonumber \\
&&
\langle
\delta^* (\eta,{\bm k}_1)
%\delta_{H}^* (\eta, {\bm k}_2) 
\delta_{x}^* (\eta, {\bm k}_2) 
\dot \delta (\eta',{\bm k}'_1)
(\delta(\eta', {\bm k}'_2)+\delta_x(\eta', {\bm k}'_2))
\rangle
A_{\ell \ell'_2  \ell''_2 1 \ell''''_2 } ^{m m'_2  m''_2 m'''_2 m''''_2}
(\eta')
[B^{m -m'_1 -m''_1}(\eta)]^*
\nonumber \\
&&
\times
j_{\ell'_1}(k_1 r) j_{\ell''_1}(k_2 r) 
{j_{\ell'_2}(k'_1 r') \over k'_1} j_{\ell''_2}(k'_2 r')
Y_{\ell'_1}^{m'_1*}(\hat {\bm k}_1) Y_{\ell''_1}^{m''_1*}( \hat {\bm k}_2)
Y^{m''''_2}_{\ell''''_2}( \hat {\bm k}'_1)
Y_{\ell''_2}^{m''_2}( \hat {\bm k}'_2),
\label{eq:pre-cross}
\end{eqnarray}
where $r'=\eta_0-\eta'$ and we use $\delta_x = -\delta_H$.

Under the assumption that all fluctuation fields are Gaussian,
the Wick theorem breaks the ensemble average in Eq.~(\ref{eq:pre-cross})
into components with $\langle \delta \delta\rangle$,
$\langle \delta_x \delta_x \rangle$ and $\langle \delta \delta_x \rangle$.
For the simplification of Eq.~(\ref{eq:pre-cross}),
we assume that $W_{21}(z)= \delta(z-z_{\rm obs})$. This is a good approximation 
because, compared to the observed frequency, the spectral resolution is narrow
(for example, the spectral resolution in the LOFAR case is less than 1 MHz 
while the observed frequency is about 150 MHz for $z_{\rm obs} \sim 10$).  
We can simplify further by using
the approximation for the integration of spherical Bessel functions
with $\ell \gg 1$,
\begin{equation}
\int dr' \int dk ~k^2 F(k) j_\ell (kr) j_\ell(kr') \approx 
\left. \int dr'{\pi \over 2} {\delta(r-r') \over r^2} F(k) \right|_{k=\ell/r}
={\pi \over 2} { F(\ell/r)\over r^2}.
\label{eq:int-bessel}
\end{equation}

Finally,
we can rewrite the cross-correlation as
\begin{eqnarray}
C^{{\rm kSZ}-21} _{\ell}
&=&-
\sum_{\ell_1  \ell_2}  
{(2 \ell _2 +1) (2 \ell_1+1) \over 2 \pi^2(2 \ell+1)}
|C^{\ell_1 \ell_2 \ell}_{000}|^2
{T_{0} (\eta_{\rm obs}) T_{\rm cmb} \over H_{\rm obs} r_{\rm obs}^2} {\dot G(\eta_{\rm obs}) 
\over G(\eta_{\rm obs})} \bar g(\eta_{\rm obs})
\int { d { k } } {j_{\ell_1}(k r_{\rm obs}) }
\left. {d j_{ \ell _1}(k r) \over d  r } \right|_{r=r_{\rm obs}}
\nonumber \\
&& \times
\left.
\left [
\left(
P_{\delta x} \left(\eta_{\rm obs}, {\ell_2 \over r} \right) 
+P_{xx} \left(\eta_{\rm obs}, {\ell_2 \over r} \right) 
\right)
P_{\delta \delta}(\eta_{\rm obs}, k)
+\left(
P_{\delta \delta} \left(\eta_{\rm obs}, {\ell_2 \over r} \right) 
+ P_{\delta x}\left(\eta ,{\ell_2 \over r} \right)
\right)
 P_{\delta x}(\eta_{\rm obs}, k)
\right] \right|_{r=r_{\rm obs}}
, 
\label{eq:cross-ksz21}
\end{eqnarray}
where the power spectra $P_{\delta \delta}, P_{x x}$ and $P_{\delta x}$
are defined as 
$\langle \delta(\eta,k_1) \delta(\eta,k_2)\rangle=(2 \pi)^3\delta(k_1-k_2)
P_{\delta \delta}(\eta,k_1)$,
$\langle \delta_x(\eta,k_1) \delta_x(\eta,k_2)\rangle=(2 \pi)^3\delta(k_1-k_2)P_{xx}(\eta,k_1)$,
and
$\langle \delta(\eta,k_1) \delta_x(\eta,k_2)\rangle=(2 \pi)^3\delta(k_1-k_2)P_{\delta x}(\eta,k_1)$.
%({\it
%We give the detail calculation in Sec.~\ref{sec:crosscal}.
%})
In Eq.~(\ref{eq:cross-ksz21}), $G$ is the growth factor of the dark matter density fluctuations
which is $\delta (k, \eta) = G(\eta) \delta(k)$ with the present density
fluctuations $\delta(k)$. Now, the epoch we are interested in is matter dominated, 
so that we can assume $G \propto 1/(1+z)$ in terms of the redshift $z$. 
%Therefore, $\dot G/G = H/(1+z)$.

%For computing the Clebsch-Gordan coefficient,
%we utilize the approximation,
%\begin{equation}
%{\left(C^{abc}_{000}\right)^2\over 2c+1} \approx {e\over 2\pi}
%\left(1+{1\over 2g}\right)^{-2g-3/2}
%\exp\left({1\over 8g}-{1\over 8(g-a)}-{1\over 8(g-b)} -
%{1\over 8(g-c)}\right) \left[g(g-a)(g-b)(g-c)\right]^{-1/2},
%\label{eq:clebsh-approx}
%\end{equation}
%where $g=(a+b+c)/2$.
%This approximation is accurate to just 1\% for the worst case 
%$a=b=c=2$.
In order to calculate the cross-correlation, the power spectra $P_{xx}$
and $P_{\delta x}$ which are determined by the reionisation model 
are essential. We discuss the analytical reionisation model in the following section.

\section{reionisation model}

For an analytical reionisation model,
we adopt the approach of \citet{furlanetto-zaldarriaga-2004}
and \citet{2005ApJ...630..643M}.  
Ionisation bubbles start to
evolve from high density galaxy regions into the voids,
as shown in recent numerical simulations \citep[e.g.][and references therein]{2009arXiv0906.4348T}.
Therefore, the mass of ionised gas $m_{\rm ion}$ is associated with the
mass of a collapsed object $m_{\rm gal}$ by the Ansatz, $m_{\rm ion}=\zeta m_{\rm gal}$ where
$\zeta$ is an ionizing efficiency. 
The condition for the full ionisation of a region of mass $m$ 
is that the region contains sufficient sources to self-ionise, 
i.e. $f_{\rm coll}\ge\zeta^{-1}$, where $f_{\rm coll}$ is the
fraction of collapsed halos above the critical mass for collapse, $m_{\rm min}$ \citep{1993MNRAS.262..627L}. 

This criterion gives the barrier (the density threshold) $\delta_x$ 
for ``self-ionisation'' which depends on $m$. 
\citet{2004ApJ...613...16F} found a reasonable approximation 
of the barrier %at the redshift $z$ 
in the linear form of the
variance of the density fluctuations, $\sigma^2(m,z)$,
% on the scale $m$ at $z$, $\sigma^2(m,z)$,
as $B(m,z)=B_0+B_1 \sigma^2(m,z)$ where $\sigma(m,z)$ is
obtained by smoothing the density field at the scale $m$. Here,
$B_0=\delta_c -\sqrt{2} K(\zeta) \sigma_{\rm min} (z)$
%$K(\zeta) = {\rm erf}^{-1}(1 - \zeta^{-1})$
and $B_1=\partial \delta_x / \partial \sigma^2 |_{\sigma^2 =0}$
where $\sigma_{\rm min} (z)$ is the mass dispersion at
the minimum mass and redshift $z$ for the collapsed ionisation source. 
  
For the linear barrier $B(m,z)$,
the bubble mass function is written as \citep{1998MNRAS.300.1057S}
%the comoving number density of HII regions with masses
\begin{equation}
\frac{d n(m)}{d m}dm=\sqrt{\frac{2}{\pi}}\frac{\bar{\rho}}{m^2}\left|\frac{d\log\sigma}{d\log m}
\right|\frac{B_0}{\sigma(m)}\exp\left[-\frac{B^2(m,z)}{2\sigma^2(m)}\right]dm\, ,
\label{eq:bubblemassfunction}
\end{equation}
where $\bar{\rho}$ is the mean mass density of the Universe.

The smallest bubble mass is given by $\zeta m_{\rm min}$. Therefore,
we can obtain the mean ionised fraction (volume averaged) $\bar x_i$ as
\begin{equation}
\bar x_i=\int_{\zeta m_{\rm min}} V(m) \frac{d n(m)}{d m}dm
=\frac{1}{2} e^{-2 B_0  B_1}  {\rm
erfc} \left( \frac{B_0 - B_1 \sigma_{\zeta}^2}{\sqrt{2  \sigma_{\zeta
}^2}} \right) 
+ \frac{1}{2} {\rm erfc} \left(\frac{B_0 + B_1 \sigma_{\zeta}^2}
{\sqrt{2  \sigma_{\zeta}^2}} \right), 
\label{eq:ionhistory}
\end{equation}
where $\sigma_{\zeta}=\sigma(\zeta m , z)$ and $V(m)$ is the comoving volume 
of a bubble with mass $m$.

In the case of a linear barrier,
the linear bias of a source of mass $m$ is given by \citep{2005ApJ...630..643M}
\begin{equation}
b(m)=1+ {B(m)/\sigma^2(m) -1/B_0 \over D(z)}.
\end{equation}
Therefore, the mean bias of the bubble $\bar b(m)$ is 
obtained from 
\begin{equation}
\bar{b} = \bar x_i^{-1} \int dm \, b(m) V(m) \frac{dn(m)}{dm}.
\label{eq:bbar}
\end{equation}

In this reionisation model, the free parameters for the model
are $\zeta$ and $m_{\rm min}$. Here we take two parameter sets
which are motivated from numerical simulations: ``stars'' model
and ``QSOs'' model \citep{2010MNRAS.tmp...65J}. 
In both models, the ionised fraction
reaches $\bar x_i= 0.5$ at $z=11$, in order to agree with the WMAP results. 

In the ``stars'' model, we assume that stars are 
responsible for reionisation. 
We take a low efficiency $\zeta \approx 40$ which is reasonable for
normal star formation  
and assume that the minimum mass corresponds to 
a virial temperature of $10^4$ K, above which cooling by atomic hydrogen becomes efficient. 

In the ``QSOs'' model, we assume that the reionisation history is faster
and the bubble size is larger compared to those in the ``stars'' model.
Therefore, we set high a virial temperature ($5 \times 10^{4}$ K) 
and a high efficiency $\zeta \approx 200$. The candidates for the ionisation sources 
are massive stars and QSOs.

We show the evolution of the ionised fraction for each model in Fig. \ref{fig:ionhistory}.
From Eq.~(\ref{eq:bubblemassfunction}), we can obtain the bubble size distribution
$V d n / dR$ as a function of the comoving size $R$ of a bubble under the assumption
that the bubbles are spherical.
We plot the results in Fig. \ref{fig:bubblesize}.

\begin{figure}
  \begin{center}
 \includegraphics[keepaspectratio=true,height=60mm]{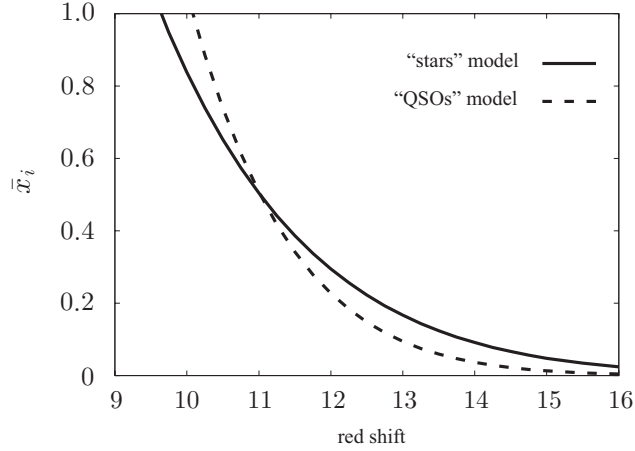}
  \end{center}
  \caption{Evolution of the mean ionised fraction. The solid and
  dotted lines represent $\bar x_i$ in the ``stars'' and ``QSOs'' models, respectively.}
  \label{fig:ionhistory}
\end{figure}

\begin{figure}
  \begin{center}
\includegraphics[keepaspectratio=true,height=60mm]{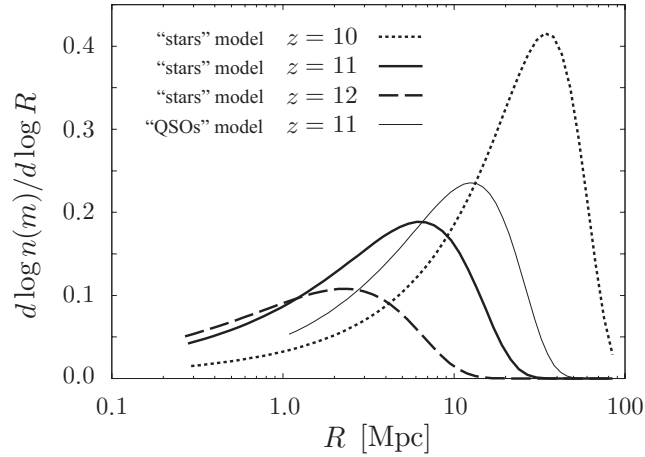}
  \end{center}
  \caption{Ionised bubble comoving size distribution. The dotted, solid and dashed
  lines represent the distributions in the ``stars'' model 
  at $z=10$, $z=11$ and $z=12$, respectively. 
  The ionised fractions are $\bar x_i=0.8$ at $z=10$, $\bar x_i=0.5$ at $z=11$
  and $\bar x_i=0.3$ at $z=12$. 
  We also plot the distribution
  in the ``QSOs'' model at $z=11$ as the thin solid line. The left side of each line
  ends at $R(\zeta m_{\rm min})$ where $\zeta m_{\rm min}$ 
is the minimum mass of the
  ionised region.}
    \label{fig:bubblesize}
\end{figure}

\subsection{The two-point correlation function $\xi_{xx} (r)$}

In order to obtain the power spectra $P_{xx}$ and $P_{\delta x}$ in Eq.~(\ref{eq:cross-ksz21}), 
we need to compute the correlation function $\xi_{xx} (r)=\langle x_i ({\bm x}_1) x_i ({\bm x}_2) 
\rangle-\bar x_i^2$ and $\xi_{\delta x} (r) = \langle \delta({\bm x}_1) x_i ({\bm x}_2) \rangle$
where the points ${\bm x}_1$ and ${\bm x}_2$ are separated by $r=|{\bm x}_1-{\bm x}_2|$. 
Here we utilize the analytical correlation functions
of \citet{2005ApJ...630..643M}.

As in the case of the density correlation function 
in the halo formalism, the correlation function of the ionised fraction $\xi_{xx} (r)$
receives two contributions. One is a one bubble term $P_1$ 
which is the two-point correlation for the case where  
two points which are separated by $r$ are ionised by the one and same ionisation source, 
the other is a two bubble term $P_2$ which corresponds to the case where
two points are ionised by two separate sources.
As shown in Fig.~\ref{fig:bubblesize}, the typical size of an ionisation bubble becomes
larger than 5 Mpc when the ionised fraction reaches one half.
In such regime, where the ionisation bubbles become large,
$P_1$ is largely dominant and $P_2$ can be ignored. 
Thus, \citet{2005ApJ...630..643M} divide the reionisation process into 
two phases: the early phase and the late phase.
In the early phase, both $P_1$ and $P_2$ are important,
while in the late phase, $P_1$ is dominant and $P_2$ can be
ignored. The criterion for these phases is set as $\bar x_i > 0.5$
in order to be in agreement with results from the hybrid
approach of analytic modeling and numerical simulations of 
\citet{2005ApJ...630..657Z}.
They define the correlation function $\xi_{xx} (r)$ by
\begin{eqnarray}
\xi_{xx} (r) & = \left \{ 
\begin{array}{l l}
(1-\bar x_i) \,P_1(r) &  {\rm when ~} \bar x_i >0.5, \\
P_1(r) + P_2(r)-\bar x_i^2  &{\rm otherwise},  
\end{array} \right.
\label{eq:xx}
\end{eqnarray}
where
\begin{eqnarray}
P_1(r) & = & \int dm \, {dn(m) \over dm}  V_0(m, r), 
\label{eq:onebub} \\
P_2(r) &=& \int dm_1 {dn(m_1) \over dm} \int d^3{\bm r}_1 
\int dm_2  {dn(m_2) \over dm} 
\int d^3{\bm r}_2 [1 + \xi({\bm r}_1 - {\bm r}_2 | m_1, m_2)].
\label{eq:twobub}
\end{eqnarray}
Here, $ \xi(r | m_1, m_2)$ is the excess probability to have 
a bubble of mass $m_1$ at the distance $r$ from a bubble of mass $m_2$.
For the simplicity of the calculation,
it is assumed that $\xi(r | m_1, m_2)$ can be written in terms of the correlation
function of the matter density $\xi_{\delta \delta}$
as $\xi(r | m_1, m_2) = \bar{b}
\xi_{\delta \delta} ({\rm max}(r, R_1 + R_2))$ 
where $R_1(m_1)$ and $R_2(m_2)$ are the bubble radii.  

In order to calculate the volume in Eqs.~(\ref{eq:onebub}) and (\ref{eq:twobub})
analytically, all ionisation bubbles are assumed spherical.
Therefore, $V_0(m, r)$ is the volume within a sphere of mass $m$ that can
encompass two points separated by a distance $r$. 
For the volume integration in Eq.~(\ref{eq:twobub}),
\citet{2005ApJ...630..643M} adopt the overlapping conditions: (1) $m_1$ cannot ionize $r_2$, 
and $m_2$ cannot ionize $r_1$; (2) the center of $m_2$ cannot lie inside $m_1$, but any
other part of $m_2$ {\it can} touch $m_1$, and vice versa. 

\subsection{The two-point cross-correlation function $\xi_{\delta x} (r)$}

As in the case of $\xi_{x x} (r)$,
the two-point cross-correlation $\xi_{\delta x} (r)$ has two contributions, $P_{\rm in}$ and $P_{\rm out}$.
The contribution $P_{\rm in}$ corresponds to the case of both points being contained within the same ionised bubble. Following \citet{2005ApJ...630..643M}, it is
written as
\begin{eqnarray}
P_{\rm in}(r) &=& \int dm {dn(m) \over dm} V_0(m, r) \int dm_h 
\frac{m_h}{\rho}
{d n_h(m_h | m) \over dm_h} \nonumber \\ 
& = & \int dm  {dn(m) \over dm}  V_0(m, r) 
[1 + B(m,z)], 
\label{eq:pin}
\end{eqnarray}
where
the last line in Eq.~(\ref{eq:pin}) is obtained by using the fact 
that the inner integral is the mean over-density of the bubble 
$1+ \delta_B $ and $\delta_B$ is $B(m,z)$ at linear order.

The contribution $P_{\rm out}$ corresponds to the case when one point
is outside the ionised bubble of the other point.
\citet{2005ApJ...630..643M} give $P_{\rm out}$ in terms of 
the mean bias for halos $\bar{b}_h$, 
\begin{eqnarray}
P_{\rm out}(r) & = & \bar{x}_i - \int dm {dn(m) \over dm}  V_0(m,r) 
+ \int dm  {dn(m) \over dm}
\int
d^3{\bm r}_ {\rm b} [\bar{b}_h \bar{b} \xi_{\delta \delta}
({\bm r} - {\bm r}_{\rm b})].
\label{eq:pout}
\end{eqnarray}
where $d n_h(m_h | m)/dm_h$ is the conditional mass function.
In Eq.~(\ref{eq:pout}), the integration range of ${\bm r}_{\rm b}$ is 
over all bubbles which ionise the point ${\bm r}_{\rm b}$ but not the other point separated 
by ${\bm r}$ from ${\bm r}_{\rm b}$.
For simplicity,  $\xi_{\delta \, \delta}$ is evaluated at the separation ${\rm max}[R(m), r ]$.

As the reionisation proceeds and the typical size of an ionised bubble
becomes large, the term $P_{out}$ becomes unimportant as compared to $P_{in}$.
Therefore, the computation of $\xi_{\delta x}$ is divided 
into two phases again,
%\citep{2005ApJ...630..643M}.
\begin{eqnarray}
\xi_{\delta x} (r) & = \left \{ \begin{array}{l l}
 P_{\rm in} - P_1 &{\rm when ~} \bar{x}_i > 0.5, \\ 
P_{\rm in} + P_{\rm out} - \bar{x}_i
& {\rm otherwise}, \end{array} \right.
\label{eq:deltax}
\end{eqnarray}
where we assume that $P_{\rm in}$ is dominant in large 
$\bar{x}_i$ ($\bar{x}_i>0.5$) and we subtract $P_1$ given by 
Eq.~(\ref{eq:pin}) from $P_{\rm in}$ which is the correlation
between $x_i$ and $\rho/ \bar{\rho}$.
%, not $x_i$ and $\delta$.

\section{Results and discussion}

We calculate the angular power spectrum of the cross-correlation described 
in Eq.~(\ref{eq:cross-ksz21})
in the two models, ``stars'' and ``QSOs''.
First, we show the results in the ``stars'' model in Fig.~\ref{fig:llcl-star.eps}.
In this model, the mean ionised fraction is $0.3$, $0.5$ and $0.8$ 
at $z=12$, $z=11$ and $z=10$, respectively.
%In our ionisation model, the power spectrum $P_{x\delta}$
%is smaller than $P_{\delta}$ at $z=10$. Since the homogeneous part dominates the bias part
%in Eq.~ in , the 21~cm and Doppler cross-correlation at $z=11$ is positive.
The signal of the cross-correlation between kSZ and 21 cm fluctuations 
exhibits an anti-correlation on small scales ($\ell > 1000$).
As mentioned in \citet{2004PhRvD..70f3509C}, 
there is a geometric cancellation in the cross-correlation.
This cancellation is responsible for a suppression of the amplitude of the cross-correlation.
However, the cross-correlation has a distinctive oscillatory shape.
Especially, we found that the peak position of the anti-correlation
represents the typical size of an ionised bubble at each redshift.
For example, at $z=11$, the typical size of an ionised bubble 
is almost $6$ Mpc, as shown in Fig.~\ref{fig:bubblesize}, 
and the anti-correlation at $z_{\rm obs}=11$ is maximal 
at the corresponding multipole $\ell \sim 4000$.
As the Universe evolves, the typical scale of an ionised bubble becomes larger.
The peak position of the anti-correlation
shifts accordingly toward smaller values. 
%Especially, the multipole where the sign of the cross-correlation changes from negative to positive
%represents the typical scale of an ionised bubble at each redshift.
%For example, at $z=11$, the typical scale of an ionised bubble 
%is almost $6$ Mpc, as shown in Fig.~\ref{fig:bubblesize}, 
%and the sign of the cross-correlation at $z_{\rm obs}=11$ changes
%at the corresponding multipole $\ell \sim 2000$.
%As the Universe evolves, the typical scale of an ionised bubble becomes larger.
%The multipole where the sign of the cross-correlation changes
%shifts accordingly toward smaller values. 

The evolution of the cross-correlation amplitude depends on the 
evolution of $\delta_x$ through the power spectra of 
$P_{xx}$ and $P_{x\delta}$ which evolve rapidly during the EoR.
Since the amplitudes of $P_{xx}$ and $P_{x\delta}$ increase as the redshift decreases,
the amplitude of the cross-correlation also becomes larger at low redshifts.
However, after the average ionisation rate reaches $\bar x_i \sim 0.9$, 
the signal of the 21 cm fluctuations
becomes weak and the cross-correlation amplitude also starts to decrease.
%
%In Fig.~\ref{fig:llcl-star.eps}, we also plot the first order cross-correlation 
%between 21~cm and CMB Doppler anisotropies
%calculated by using the same expression as Eq.~(15) of \citet{2006ApJ...647..840A}. 
%The sign of the first order cross-correlation depends on the evolution of $\delta_x$.
%As long as $\delta_x$ is small, the ionisation process is homogeneous, and 
%the cross-correlation is negative. On the other hand, in the case of a highly inhomogeneous reionisation, 
%the sign of the cross-correlation is positive.
%In our reionisation model, the first order cross-correlation at $z_{\rm obs}=11$ is negative. 
%The amplitude is $300~\mu {\rm K}^2$ at the peak position, $\ell \sim 100$, and decreases rapidly towards zero 
%at large multipoles. As we can see, the second order kSZ-21 cm cross-correlation dominates the first order cross-correlation
%at multipoles larger than $\ell = 1000$.
%However, at $z=10$, the ionisation fraction is highly inhomogeneous and $\delta_x$ 
%is evolved well. As a result, the first order cross-correlation has a positive sign
%and high amplitude. The first order cross-correlation becomes comparable to the second order kSZ-21~cm 
%cross-correlation even at $\ell \sim 1000$, while
%the kSZ cross-correlation still dominate the first order cross-correlation
%at multipoles higher than $\ell =1000$. 
%

In Fig.~\ref{fig:llcl-star.eps}, we also plot the first order cross-correlation 
between 21~cm and CMB Doppler anisotropies
calculated by using the same expression as Eq.~(15) of \citet{2006ApJ...647..840A}. 
The sign of the first order cross-correlation depends on the evolution of $\delta_x$.
As long as $\delta_x$ is small, the ionisation process is homogeneous, and 
the cross-correlation is negative. On the other hand, in the case of a highly inhomogeneous reionisation, 
the sign of the cross-correlation is positive.
In our reionisation model, the first order cross-correlation at the early phase
of reionisation is negative at $\ell < 1000$ (see the top and middle panels in 
Fig.~\ref{fig:llcl-star.eps}). We found the amplitude of the first order 
cross-correlation at $z_{\rm obs}=11$ is $300~\mu {\rm K}^2$ at the peak
position, $\ell \sim 100$, and decreases rapidly towards zero at large 
multipoles. As we can see in Fig.~\ref{fig:llcl-star.eps}, the second order 
kSZ-21 cm cross-correlation dominates the first order cross-correlation at 
multipoles larger than $\ell = 1000$.
However, as the ionisation process proceeds, the ionisation fraction is highly
inhomogeneous and $\delta_x$ is evolved well. As a result, the first order 
cross-correlation has a positive sign and a high amplitude as shown in the bottom
panel of Fig.~\ref{fig:llcl-star.eps}. The first order cross-correlation 
becomes comparable to the second order kSZ-21~cm cross-correlation even at 
$\ell \sim 1000$, while the kSZ cross-correlation still dominate the first 
order cross-correlation and has negative correlation at multipoles higher 
than $\ell =1000$.

\begin{figure}
  \begin{center}
    \includegraphics[keepaspectratio=true,height=150mm]{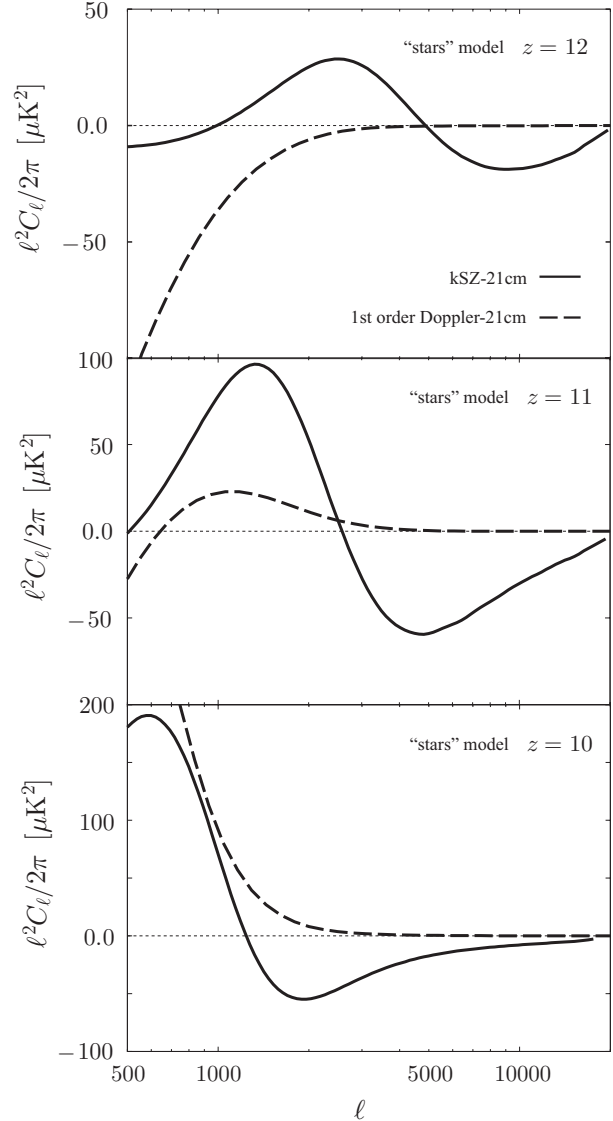}
  \end{center}
% \caption{Angular power spectra of the second order cross-correlation
% in the ``stars'' model. We plot the angular power spectra at 
% $z_{\rm obs}=10$, $z_{\rm obs}=11$ and $z_{\rm obs}=12$ as the
% dotted, solid, and dashed lines, respectively. The mean ionised 
% fraction is $\bar x =0.8$ at $z =10$, $\bar x =0.5$ at $z=11$ and
% $\bar x =0.3$ at $z=12$. For reference, we show the first order 
% cross-correlation between the CMB temperature and 21~cm
% fluctuations at $z=10$ as the thin solid line.}
 \caption{Angular power spectra of the second order cross-correlation
 in the ``stars'' model. From top to bottom panels, we plot the angular power spectra at $z_{\rm obs}=12$, $z_{\rm obs}=11$ and 
 $z_{\rm obs}=10$, respectively. The mean ionised 
 fraction is $\bar x =0.3$ at $z =12$, $\bar x =0.5$ at $z=11$ and
 $\bar x =0.8$ at $z=10$. For reference, we show the first order 
 cross-correlation between the CMB temperature and 21~cm fluctuations
 as the dotted line in each panel.}

  \label{fig:llcl-star.eps}
\end{figure}

Next we show the dependence of the angular cross-correlation power spectrum on
the ionisation model in Fig.~\ref{fig:llcl-qso.eps}.
In the ``QSOs'' model, the ionisation history is rapid and the typical size
of ionised bubbles is large.
The amplitude of $P_{xx}$ and $P_{x\delta}$ in the ``QSOs'' model
is larger than in the ``stars'' model. 
As a result, in the ``QSOs'' model, the signal of the 
cross-correlation is large and 
%the sign change 
the peak position of the anti-correlation appears on small 
multipoles, as expected.
We can therefore conclude that the cross-correlation between kSZ and 21 cm fluctuations 
at the second order is sensitive to the average size of an ionised bubble.
The first order cross-correlation also has a higher amplitude than in the ``stars'' model
because the amplitude depends on the evolution rate of the background ionisation fraction.
However, the inhomogeneous contribution coming from the term with $P_{x\delta}$ in
Eq.~(15) in \citet{2006ApJ...647..840A}
is partially canceled by the homogeneous one from the term with $P_{\delta \delta}$. As a result,
in the highly inhomogeneous ``QSOs'' reionisation model,
the cross-correlation between kSZ and second order 21 cm fluctuations
reaches a significant amplitude, compared with
the first order cross-correlation at small scales ($\ell \lesssim 1000$).
%
%{
%\it I am afraid that the part in bold style mislead a reader into 
%considering that the kSZ-21cm cross-correlation becomes comparable 
%with the 1st order cross-correlation on scales where the kSZ-21cm
%cross-correlation is positive (\ell \sim 1000) in all rapid 
%reionisation model. 
%}

\begin{figure}
  \begin{center}
    \includegraphics[keepaspectratio=true,height=60mm]{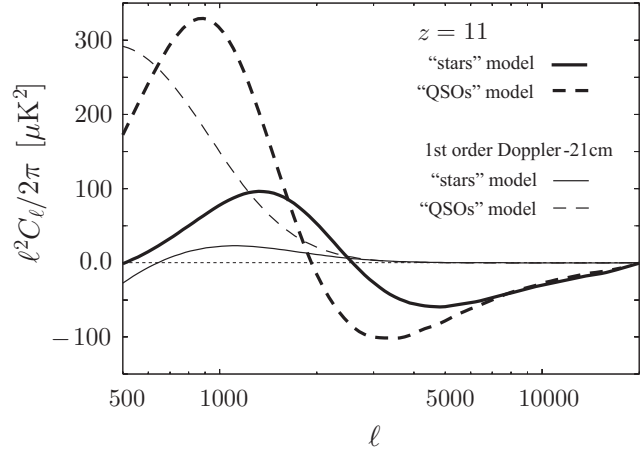}
  \end{center}
  \caption{Dependence of the cross-correlation on the ionisation model.
  The solid and the dashed lines are the power spectrum for the ``stars'' model
  and the ``QSOs'' model, respectively (see text). We set $z_{\rm obs}=11$ where $\bar x=0.5$ in
  both models. For reference, we plot the first order cross-correlation in each
  model as the thin lines.}
  \label{fig:llcl-qso.eps}
\end{figure}

\subsection{Detectability}

In the previous section, we showed that the peak position of the 
anti-correlation is related to the typical bubble size at the 
observed redshift of 21 cm fluctuations.
Here, our concern is the detectability of such negative 
peak in the kSZ-21 cm cross-correlation.
In order to investigate the detectability, we calculate the signal-to-noise ratio ($S/N$). 
For simplicity, we assume that CMB, 21 cm fluctuations and
instrumental noise are Gaussian.
The total $S/N$ can be calculated as
\begin{equation}
\left( {S \over N} \right) ^2 =
f_{\rm sky} \sum_{\ell = \ell_{\rm min}} ^{\ell_{\rm max}} (2 \ell +1)
{| C_\ell ^{21-{\rm CMB}}| ^2  \over | C_\ell ^{21-{\rm CMB}} |^2
+C_\ell ^{21} C_\ell ^{\rm CMB}},
\label{eq:SNratio}
\end{equation}
where $f_{\rm sky}$ is the sky fraction common to the two 
cross-correlated signals, and $C_\ell ^{\rm CMB}$, $C_\ell ^{21}$ and
$C_\ell^{21-{\rm CMB}}$ are the angular power spectra of CMB, 21 cm 
fluctuation and the cross-correlation between 21 cm and CMB,
respectively.
In order to focus on the detectability of the signal from the typical
bubble size, we set $\ell_{\rm min}=500$ and $\ell_{\rm max}=5000$.

At the multipoles that we are interested in ($\ell>1000$), 
the dominant CMB signal is due to the thermal 
SZ effect \citep{1969Ap&SS...4..301Z}. However we can remove this
contribution because of the frequency dependence of the SZ effect.
Therefore with the assumption that the foreground can be completely
removed from the CMB map, the main contribution to $C_l^{\rm CMB}$
comes from the primordial CMB anisotropies $C_\ell ^{\rm pri}$ and 
the noise of the instrument $N_\ell^{\rm CMB}$. We can write 
$C_\ell^{\rm CMB}$ as
\begin{equation}
C_\ell^{\rm CMB} = C_\ell ^{\rm pri} 
\exp (-\ell^2 \sigma_{ \rm CMB}^2/2 ) + N_\ell^{\rm CMB},
\label{eq:cmb-signal}
\end{equation}
where we assume the beam profile of CMB observation is Gaussian with
the Full Width at Half Maximum of the beam $\theta_{\rm CMB}$, and
$\sigma_{ \rm CMB} = \theta_{\rm CMB}/\sqrt{8 \ln 2}$.
The effect of the beam size is a damping of the signal of the primordial
CMB on smaller scales than the FWHM.
The noise power spectrum $N_\ell ^{\rm CMB}$ is given by 
\citep{knox-1995}
\begin{equation}
N_\ell ^{\rm CMB}=  {\sigma_{\rm pix}^2 \Omega_{\rm pix}},
\end{equation}
where $\sigma_{\rm pix}$ is the sensitivity in each pixel and 
$\Omega_{\rm pix}$ is the solid angle per pixel;
$\Omega_{\rm pix}= \theta_{\rm CMB}^2$. 

As for the 21~cm fluctuations, the noise signal from the instruments
and foreground will dominate the intrinsic signal from the EoR.
Assuming that the foreground can be removed to the level below the
noise from instruments, we can write 
\begin{equation}
C_\ell ^{21} =N_\ell^{21}=
{2 \pi \over t_{\rm obs} \Delta \nu} \left( {D \lambda \over A/T} 
\right)^2,
\label{eq:21instnoise}
\end{equation}
where we use the noise power spectrum of 21~cm observation 
estimated by \citet{2004ApJ...608..622Z}. 
In Eq.~(\ref{eq:21instnoise}),
$\Delta \nu$ is the bandwidth, $t_{\rm obs}$ is the total
integration time, $A/T$ is the sensitivity (an effective area divided
by the system temperature) and $D$ is the length of the baseline associated with the FWHM of the 21~cm observation 
$\theta_{21} =\lambda /D $.

In the calculation of the cross-correlation signal, we assume that 
the foregrounds and noise of 21~cm fluctuations and CMB anisotropy 
are not correlated. Therefore, the cross-correlation consists mainly 
of the first order Doppler-21~cm cross-correlation and 
the second order kSZ-21~cm one,
\begin{equation}
|C_\ell ^{21-{\rm CMB}}| ^2= (|C_\ell ^{21-{\rm Dopper}}|^2 +
|C_\ell ^{21-{\rm kSZ}}|^2) \exp [-\ell^2 (\sigma_{ \rm CMB}^2
+\sigma_{21}^2 )/2],
\end{equation}
where $\sigma_{21} = \theta_{21}/\sqrt{8 \ln 2}$ and
both signals are affected by the angular resolution of the
observations.

Our interest is the detectability of the cross-correlation signal
from the patchy reionisation by {\it Planck} and SKA.
Therefore, in the computation of Eq.~(\ref{eq:cmb-signal}), we adopt 
the typical value of {\it Planck} which are 
$\theta_{\rm CMB}= 5~$arcmin 
and $\sigma_{\rm pix}= 5 \times 10^{-6}$.
The goal sensitivity of SKA is currently designed 
as $A/T=5000~{\rm m^2 K^{-1}}$ at 200 MHz.
The configuration area is 20 \% of total collecting area 
for 1 km baseline, 50 \% for 5 km baseline, 75 \% 
for 150 km baseline. Because we are interested in the scales 
$\ell \sim 2000$, we take $D=1~$km and $A/T=1000~{\rm m^2 K^{-1}}$.
The sky fraction $f_{\rm sky}$ corresponds to the one of 
SKA because we consider {\it Planck} as CMB observation, which is
almost full-sky.
We assume $200 {\rm deg}^2$ per field of view and 4 independent 
survey fields for SKA. Therefore the total 
sky fraction is $f_{\rm sky} \sim 0.02$.

We plot $S/N$ as a function of $t_{obs}$ in units of 
hours for ``stars'' and ``QSOs'' models at $z=11$ in 
Fig.~\ref{fig:snratio}.  In both panels in Fig.~\ref{fig:snratio}, 
$S/N$ of SKA with {\it Planck} is represented by the solid lines.
Obviously, longer observation times make $S/N$ larger. Then,
since the cross-correlation amplitude in the``QSOs'' model is
higher than in the ``stars'' model, $S/N$ in the former model
is lager than in the later model. However both $S/N$ are below 
the detection level.
This difficulty of the detection is mainly due to the instrumental
noise of the 21 cm observation.
Although the primary CMB is one of the significant sources of
noise in the detection of the cross-correlation signal between CMB 
and 21 cm from EoR on large scales \citep{2010MNRAS.tmp...65J, 2010MNRAS.402.2617T}, the primary CMB suffers Silk damping on the
scales we are interested in here and the noise of {\it Planck} is
also kept below the sufficient level.

In order to clarify the impact of the improvement in the sensitivity
of 21~cm observation, we calculate $S/N$ in the case of a 5
times better sensitivity than that of SKA and plot the result as the
dotted line in Fig.~\ref{fig:snratio}. The improvement of the
sensitivity of 21~cm observation brings large $S/N$. Especially, 
the $S/N$ in the ``QSOs'' model can reach $S/N \sim 5$ in 500-hour
observation.
Finally, while we use the same sky fraction $f_{\rm sky} \sim 0.02$ 
in all calculations, larger sky fractions also make
$S/N$ higher. %According to Eq.~(\ref{eq:SNratio}), the sky fraction
%scales $S/N$ by $\sqrt{f_{\rm sky}}$. Therefore, in addition to
%the improvement of the sensitivity of 21~cm observations, taking
%the large survey area is effective in order to obtain large $S/N$.

\begin{figure}
  \begin{tabular}{cc}
   \begin{minipage}{0.5\textwidth}
  \begin{center}
\includegraphics[keepaspectratio=true,height=50mm]{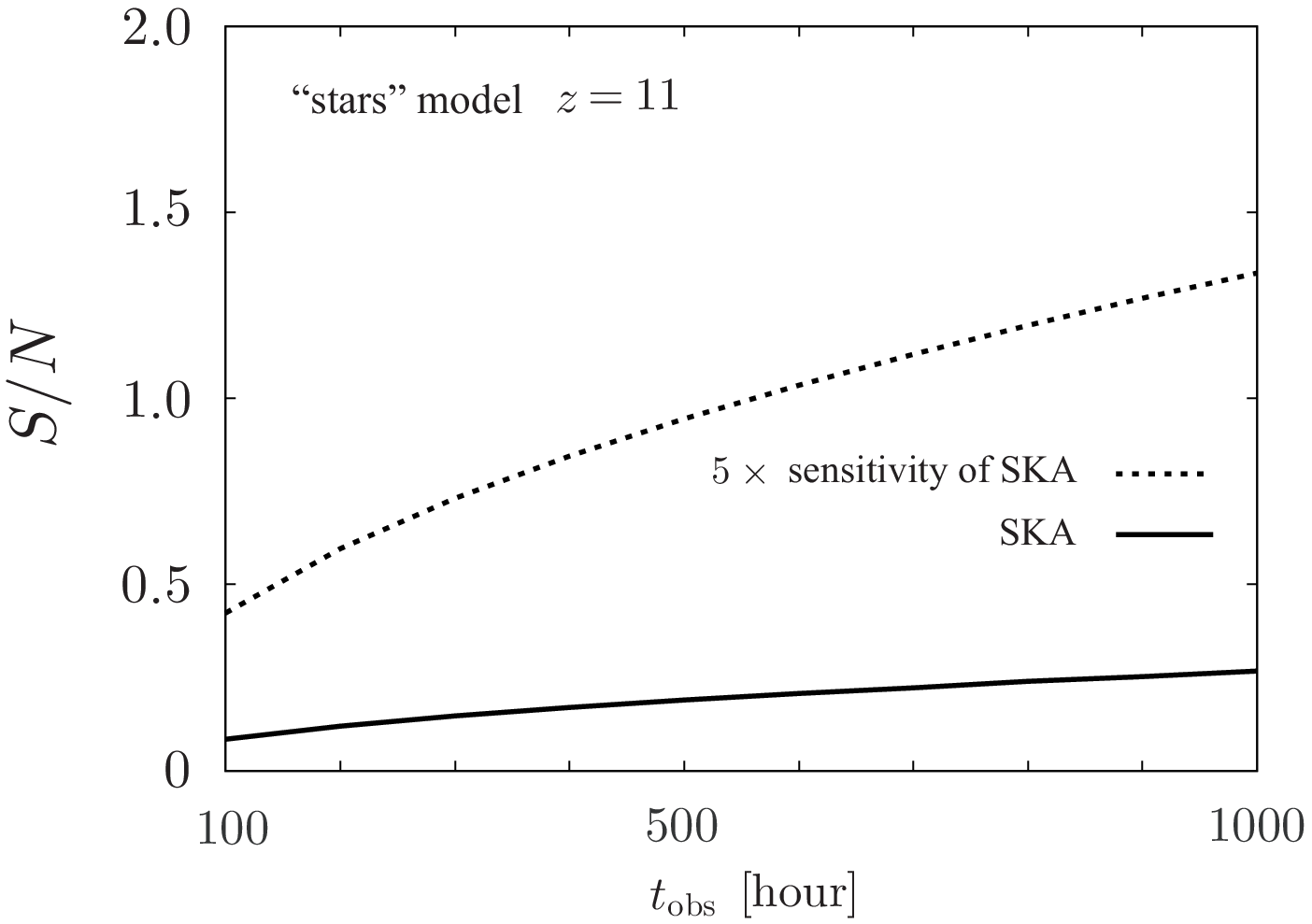}
  \end{center}
  \end{minipage}
   \begin{minipage}{0.5\textwidth}
  \begin{center}
\includegraphics[keepaspectratio=true,height=50mm]{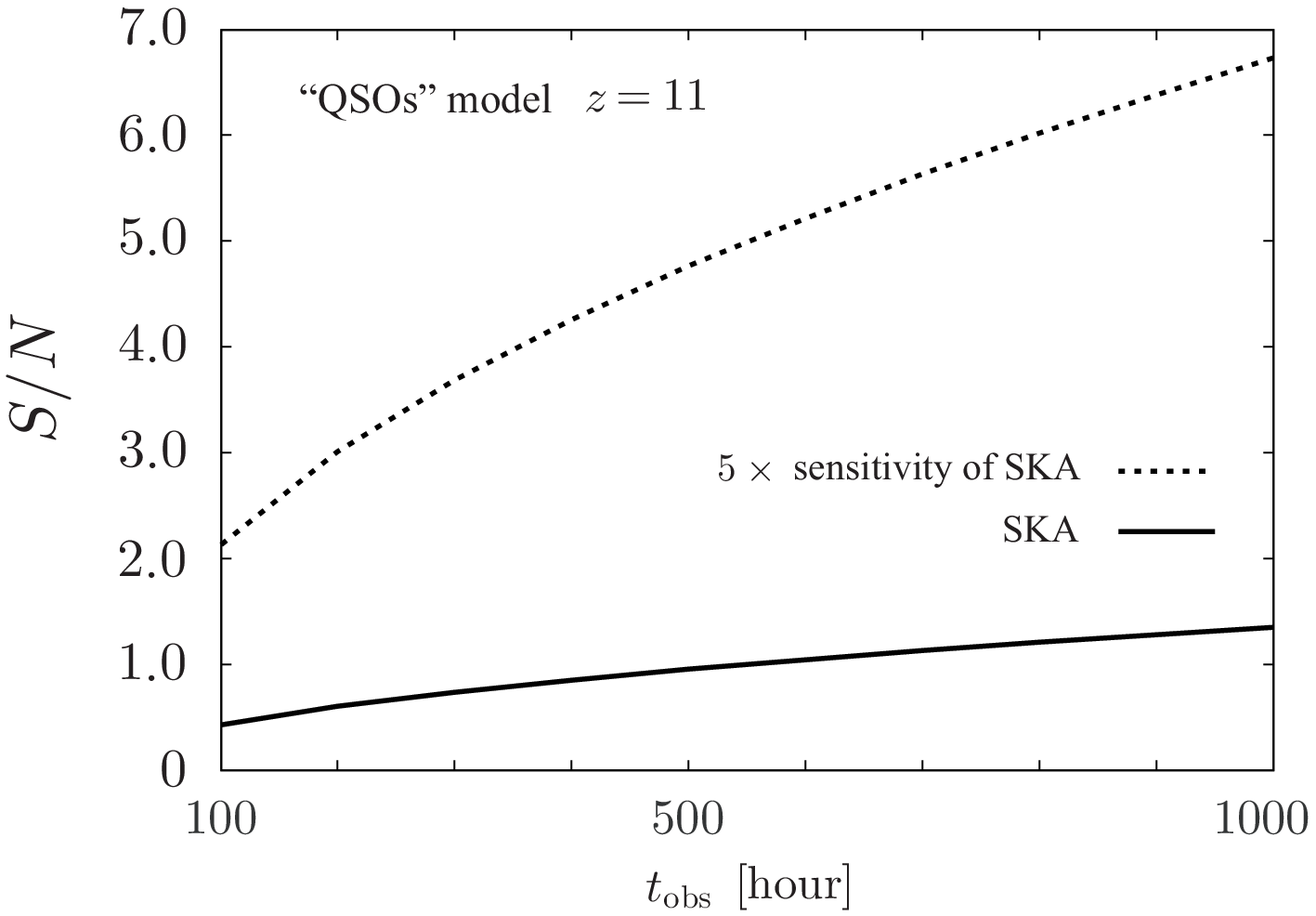}
  \end{center}
   \end{minipage}
  \end{tabular}
  \caption{The $S/N$ ratio for the detection of the cross-correlation
signal at $z=11$ as a function of the observation time. The left 
panel is for the ``stars'' model and the right panel is for the 
``QSOs'' model. In both panels, the solid and dotted lines represent
$S/N$ for SKA and for the observation with a 5
times better sensitivity than that of SKA, respectively. 
We set $f_{\rm sky} \sim 0.02$ in all plots.
}
  \label{fig:snratio}
\end{figure}

\section{conclusion}

We investigated the small scale cross-correlation between CMB anisotropies
and the 21~cm fluctuations during the EoR in harmonic space.
The CMB anisotropies at small scales are mainly caused by the kSZ effect which
is the second order fluctuation effect generated by the peculiar velocity and 
the fluctuations of the visibility
function. We therefore calculated the cross-correlation with the second order
fluctuations of 21~cm fluctuations.

The cross-correlation signal between kSZ and 21~cm fluctuations is negative on small scales. 
This anti-correlation on small scales was found in the numerical
simulations of \citet{2005MNRAS.360.1063S} and 
\citet{2010MNRAS.tmp...65J}.
%Both simulation studies showed that the signals of CMB kSZ and 21~cm
%fluctuations were anti-correlated on smaller scales than the typical
%size of a bubble.
We found that the position of the negative peak is
at the angular scale corresponding to
the typical size of an ionised bubble at the redshift probed by 21~cm fluctuation measurements.
This angular scale shifts to larger scales as ionised bubbles evolve.
The amplitude also increases with the reionisation process until
the average ionisation fraction reaches $\bar x_i \sim 0.9$.
The amplitude of the cross-correlation strongly depends on the typical bubble size. 
The cross-correlation in the case of larger bubbles 
has a higher amplitude than in the case of smaller bubbles, 
even if in both cases the mean ionisation fractions are the same.
Moreover, the amplitude of the cross-correlation from large ionised bubbles
is comparable to that of the first order cross-correlation.
 Those characteristic features of the cross-correlation could be used to distinguish 
between different reionisation histories with future observations.

We also estimated the detectability of the small-scale cross-correlation by
the current design sensitivity of SKA. It is rather difficult
%, but in principle not impossible, 
to detect the cross-correlation
signal even in the radical reionisation cases. However, if the sensitivity is
improved by a factor of 5, the detection or non-detection of the cross-correlation
signal will definitely provide information about the EoR.

\section*{Acknowledgments}
HT is supported by the Belgian Federal Office for Scientific,
Technical and Cultural Affairs through the Interuniversity Attraction Pole P6/11.

%\bibliography{ksz21cm.bib}

\end{document}